\documentclass[%
twocolumn,
 amsmath,amssymb,
 aps,
pra,
superscriptaddress,
]{revtex4-1}
\usepackage{graphicx}
\usepackage{dcolumn}
\usepackage{graphicx}
\usepackage{amsmath}
\usepackage{amsthm}
\usepackage{bm}
\usepackage{layout}
\usepackage{float}
\usepackage{amsfonts}
\usepackage{amssymb}%
\usepackage[margin=1in]{geometry}
\usepackage{bbm}
\usepackage[export]{adjustbox}

\newcommand{\ket}[1]{\left | #1 \right\rangle}
\newcommand{\bra}[1]{\left \langle #1 \right |}

\renewcommand{\epsilon}{\varepsilon}

\usepackage{color}

\begin{document}


\title{Quantum simulation of ferromagnetic Heisenberg model}

\author{Yiping~Wang}
\affiliation{Joint Center for Quantum Information and Computer Science, NIST/University of Maryland, College Park, Maryland 20742, USA}
\affiliation{Joint Quantum Institute, NIST/University of Maryland, College Park, Maryland 20742, USA}
\affiliation{School of Physics, Nanjing University, Nanjing 210093, China}

\author{Minh~Cong~Tran}
\affiliation{Joint Center for Quantum Information and Computer Science, NIST/University of Maryland, College Park, Maryland 20742, USA}
\affiliation{Joint Quantum Institute, NIST/University of Maryland, College Park, Maryland 20742, USA}

\author{Jacob~M.~Taylor}
\affiliation{Joint Center for Quantum Information and Computer Science, NIST/University of Maryland, College Park, Maryland 20742, USA}
\affiliation{Joint Quantum Institute, NIST/University of Maryland, College Park, Maryland 20742, USA}
\affiliation{Research Center for Advanced Science and Technology,
University of Tokyo, Meguro-ku, Tokyo 153-8904, Japan}

\begin{abstract}
Large quantum simulators, with sufficiently many qubits to be impossible to simulate classically, become hard to experimentally validate.
We propose two tests of a quantum simulator with Heisenberg interaction in a linear chain of spins. 
In the first, we propagate half of a singlet state through a chain of spin with a ferromagnetic interaction and subsequently recover the state with an antiferromagnetic interaction. 
The antiferromagnetic interaction is intrinsic to the system while the ferromagnetic one can be simulated by a sequence of time-dependent controls of the antiferromagnetic interaction and Suzuki-Trotter approximations.
In the second test, we use the same technique to transfer a spin singlet state from one end of a spin chain to the other. 
We show that the tests are robust against parametric errors in operation of the simulator and may be applicable even without error correction.
\end{abstract}

\pacs{Valid PACS appear here}
\maketitle


\section{Introduction}
Quantum simulators~\cite{feynman82} provide the opportunity for a controlled quantum system to emulate the behavior of another system whose properties we would like to better understand. Progress towards building quantum simulators is occurring rapidly, with demonstrations up to and beyond 50 qubits~\cite{houck12,ma14,omalley16,hensgens17,loredo16,monroe17,lukin17}. 
However, a key question remains: how do we test the behavior of such a system without fault tolerance while at the same time dealing with the exponential growth of classical simulation costs~\cite{vazirani14,nori14}?  
A variety of approaches are being considered in this domain, including comparison of classical versus quantum behavior~\cite{albash2015,kafri2015,hangleiter2017}
or demonstrating so-called~\cite{wiesner17} 
`quantum supremacy'~\cite{calude17,neill17,miller2017,bermejo2017,aaronson2016,bremner2016,boixo2016}.

Here we consider an approach for testing the performance of a spin-based quantum simulator that can be easily implemented in quantum dot computing systems~\cite{Hanson17,hensgens17,gray16}, 
as well as other systems that have nearest neighbor Heisenberg interactions~\cite{porras2004,salath15,grass2014,ma2011}.
Starting with an intrinsically antiferromagnetic system, we show how time-dependent control of the exchange interaction enables us to make a Suzuki-Trotter-type simulation of a ferromagnetic system. This in turn allows us to propose two different tests for a linear chain of spins. In the first, one does a Loschmidt echo, propagating a single up spin through a chain of down spins with the ferromagnetic interaction, then back with the antiferromagnetic interaction. In the second, one transfers a spin singlet through a chain of spins following the protocol outlined in Ref.~\cite{StateTransfer}. Successful recovery of the singlet on the far end provides a test of the quantum channel capacity of the underlying quantum simulator. These techniques are ideal for quantum dot-based computer, where preparation and measurement of singlet states~\cite{petta05,taylor07} 
and antiferromagnetic Heisenberg interactions~\cite{loss98} 
are natural elements of the system.

\section{Simulating ferromagnet with antiferromagnet}
 In this section, we describe how to simulate time evolution under a ferromagnetic interaction using an antiferromagnetic interaction. In our scenario, we assume that one can prepare $n$ spins and let them evolve for some time $t$ under antiferromagnetic nearest-neighbor interactions:
\begin{align}
	H_{a,n} = \sum_{i=1}^{n-1} J_{i,i+1}(t) \bm S_i\cdot \bm S_{i+1}, \label{EQ_Ha}
\end{align}
where $\bm S_{i}$ is the spin operator vector of the $i$th qubit and the arbitrary positive $J_{i,i+1}(t)$ are tunable parameters. In what follows, we take $\hbar =1$.
Using the above Hamiltonian, we would like to simulate a ferromagnetic Hamiltonian,
\begin{align}
	H_{f,n} = -\sum_{i=1}^{n-1} \tilde J_{i,i+1}(t) \bm S_i\cdot \bm S_{i+1}, \label{EQ_Hf}
\end{align}
for arbitrary positive parameters $\tilde J_{i,i+1}(t)$. Let us first consider a two-spin system, i.e. $n = 2$.
\subsection{Basic element: a two-spin system}
\label{2a}
In this case, the antiferromagnetic and ferromagnetic Hamiltonians in Eq.~\eqref{EQ_Ha} and Eq.~\eqref{EQ_Hf} are reduced to 
\begin{align}
	H_{a,2} &=  J_{12}(t) \bm S_1\cdot \bm S_2, \\
    H_{f,2} &= -\tilde J_{12}(t) \bm S_1\cdot \bm S_2.\label{2}
\end{align}
Let us first consider time-independent Hamiltonians, i.e. $J_1(t) = J$ and $\tilde J_1(t) = \tilde J$, for all $t$.

In order to investigate the simulation of time evolution under the ferromagnetic and the antiferromagnetic Hamiltonians (Eq.~\eqref{EQ_Ha},~\eqref{EQ_Hf}), we first prepare an arbitrary two-spin initial state $\ket{\psi\left(0\right)}$ at $t=0$. 
Let it evolve under the ferromagnetic interaction $H_{f,2}$ in Eq.~\eqref{2} for time $ t $ and the time evolution operator would be $ U_f(t)=e^{-iH_{f,2}t} $ . Since we can always represent the state $\ket{\psi(0)}$ in the eigenstates of $H_{f,2}$, i.e.
\begin{align}
\ket{\psi(0)}=c_0\ket{s}+\sum_{m=1}^{3} c_m\ket{t_m},
\end{align}
where $c_0$ and $c_m$ are coefficients,
the state of the system at time $t$ will be 
\begin{align}
\ket{\tilde\psi\left(t\right)}&=U_f(t)\ket{\psi\left(0\right)}\nonumber\\
\label{e1}
&=c_0e^{i\tilde J\epsilon_st}\ket{s}+e^{i\tilde J\epsilon_tt}\sum_{m=1}^{3} c_m\ket{t_m}\\
\label{e2}
&=e^{i\tilde J\epsilon_tt}\left(c_0e^{i\tilde J\Delta\epsilon t}\ket{s}+\sum_{m=1}^{3} c_m\ket{t_m}\right),
\end{align}
with
\begin{align}
	\Delta\epsilon=\epsilon_s-\epsilon_t,
\end{align}
where $ \epsilon_t=\frac{1}{4} $ and $\epsilon_s=-\frac{3}{4} $ are the eigenvalues of two-spin system, the triplets  $ \ket{t_m} $ and the singlet $\ket{s}$ respectively.  The degeneracy of the triplets allows us to simplify Eq.~\eqref{e1} to~\eqref{e2}. Similarly, for the antiferromagnetic interaction, $ U_{af}(t')=e^{-iH_{a,2}t'} $, we have
\begin{align}
\label{e3}
\ket{{\psi}(t')}=e^{-iJ\epsilon_tt'}\left(c_0e^{-iJ\Delta\epsilon t'}\ket{s}+\sum_{m=1}^{3} c_m\ket{t_m}\right).\end{align}

\begin{figure}[t]
	\includegraphics[width=0.4\textwidth]{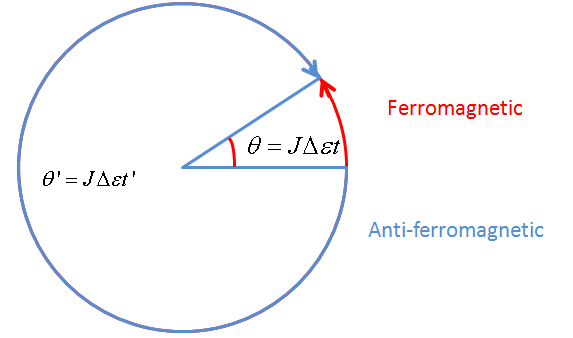}
	\caption{The rotation angles $ \theta $ and $ \theta' $ generated by ferromagnetic (red) and antiferromagnetic (blue) interaction are of opposite clockwise direction. To let these two rotations end up with the same effect, there should be $ \theta=2\pi-\theta' $. We show how this allows us to simulate ferromagnet with antiferromagnet in section \ref{2a}.}
	\label{trick}
\end{figure}

In this two-spin case, we can describe the evolution by two phase terms. The first terms in Eq.~\eqref{e2},~\eqref{e3} are global phases. Meanwhile, the phase added to the singlet will cause the twist of the spin chain on which we lay more focus.  Also, these evolutions as stated in Eq.~\eqref{e1} and Eq.~\eqref{e3} are periodic and their periods are correlated to $ J $ and $\tilde{J}$. Now we try to utilize this periodic property to realize our simulation of the ferromagnetic interaction by the antiferromagnetic one. The goal here is to find $t$ and $t'$ for the ferromagnetic and the antiferromagnetic time evolutions respectively which will bring about identical final states:
\begin{align}
	\ket{\tilde\psi(t)}=\ket{{\psi}(t')},\label{psi}
\end{align}
up to a possibly a global phase.
Here we state how evolution of different $t$ can make our desired simulation possible. Due to the fact that $e^{i\theta(t)}=e^{-i(2k\pi-\theta(t))}=e^{-i\theta'(t')}$, where $\theta(t)$ and $\theta'(t')$ refer to the phases in $\ket{\tilde\psi(t)}$ and $\ket{\psi(t')}$, we wish to find a proper relation between $t$ and $t'$ which will make Eq.~\eqref{psi} possible (Fig.~\ref{trick}). The rotation angle $ \theta(t)=\tilde J\Delta\epsilon t $ generated by the ferromagnetic interaction is anti-clockwise while the angle $ \theta'(t')=J\Delta\epsilon t' $ generated by the antiferromagnetic interaction is clockwise. A restriction for $ t $ and $ t' $ exists to enable $ \theta=2\pi-\theta' $. To translate these into equations, the times $t$ and $t'$ must satisfy
\begin{align}
	\tilde J\Delta\epsilon t&=2k\pi-J\Delta\epsilon t',
\end{align}
for some integers $k$. Solving the equation, we have the relation between $ t $ and $ t' $
\begin{align}
\label{t'}
t'=\frac{\tilde J}{J}\left(\frac{2\pi}{\tilde J|\Delta \epsilon|}-t\right)\ \  \ (k=-1)
.\end{align}
The value of $k$ are chosen to give a minimal experimental time $t'$.
According to this restriction, when we start with the same initial state, an evolution for $ t' $ under an antiferromagnetic interaction is equivalent to an time evolution for $ t $ under a ferromagnetic interaction. That is
\begin{align}
\label{T}
U_f\left(t\right)\ket{\psi}=U_{af}\left(t'\right)\ket{\psi},
\end{align}
up to a global phase for any two-spin states $\ket{\psi}$.

Note that while our discussion is for a time-independent interaction, we can also use this technique to simulate a time-dependent $J$ by splitting into smaller time periods in each of which we assume $J$ to be a constant.
\subsection{Trotterization for a larger spin chain}

In order to simulate the Hamiltonian in Eq.~\eqref{EQ_Hf}, we start from simulating for three-spin case, where the contributing Hamiltonians are
\begin{align}
&H_{12}=-J_{1,2}\bm S_1\cdot\bm S_2,\\
&H_{23}=-J_{2,3}\bm S_2\cdot\bm S_3.
\end{align}
Using the protocol from the last subsection, we can simulate $U_{12}(t)=e^{-iH_{12}t}$ and $U_{23}(t)=e^{-iH_{23}t}$. However, since $H_{12}$ and $H_{23}$ do not commute, i.e. $ \left[H_{12},H_{23}\right]\neq0 $, a direct combination of these two time evolution operations is not equivalent to the system we intend to simulate:
\begin{align}
e^{-i(H_{12}+H_{23})t}\neq e^{-iH_{12}t}e^{-iH_{23}t}.
\end{align}

Instead, we use a Trotterization technique~\cite{trotter59}. The Trotter formula is a good way to approximate the time evolution Hamiltonian $H_{f,3}=H_{12}+H_{23}$, with the two-body interaction Hamiltonians. Here we use second-order Trotter expansion which gives a considerably small error term:
\begin{align}
e^{-iHt}=\left(e^{-i\frac{H_{12}}{2}\frac{t}{N}}e^{-iH_{23}\frac{t}{N}}e^{-i\frac{H_{12}}{2}\frac{t}{N}}\right)^N+O\left(\frac{t^3}{N^2}\right),
\end{align}
where $N$ is the number of  Trotter steps which can be increased to reduce the approximation error for a given time $t$. Therefore, the time evolution $e^{-iH_{f,2}t}$ can be approximated by a series of alternative time evolutions under $H_{12}$ and $H_{23}$, each of which can be simulated using our technique described in II.A. 

\begin{figure}[t]
	\centering
	\includegraphics[scale=0.4]{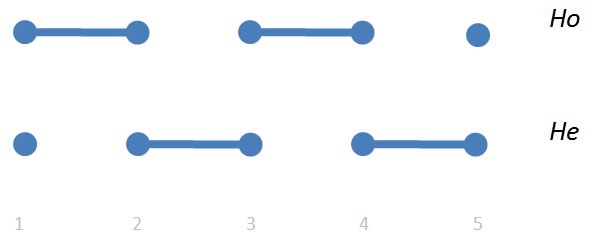}
	\caption{Illustration of the interactions in $ H_o $ and $ H_e $ in Eq.~\eqref{ho} and Eq.~\eqref{he}. The dots represent spins and the links between them refer to the included interactions.}
	\label{eo}
\end{figure}

Furthermore, we can use the same technique to simulate time evolution under the general Hamiltonian in Eq.~\eqref{EQ_Hf} with $n$ spins. To do that, we group the terms in Eq.~\eqref{EQ_Hf} in terms of time evolution under $H_o$ and $H_e$ (Fig.~\ref{eo})
\begin{align}
\label{ho}
 H_{o}&=J_{1,2}\bm{S}_1\cdot\bm{S}_2+J_{3,4}\bm{S}_3\cdot\bm{S}_4+\dots, \\
\label{he}
H_{e}&=J_{2,3}\bm{S}_2\cdot\bm{S}_3+J_{4,5}\bm{S}_4\cdot\bm{S}_5+\dots,
\end{align}
where $H_{o}+H_{e}=-H_{f,n}$. The terms in Eq.~\eqref{ho} (Eq.~\eqref{he}) mutually commute with each other. Therefore, we can further expand the time evolution under $H_{o}(H_{e})$ in terms of time evolutions under each interaction pair and simulate those within a same group simultaneously, i.e., 
\begin{align}
e^{-iH_{o}t'}=e^{-iH_{12}t'}e^{-iH_{34}t'}\dots.
\end{align}
And with our technique in section \ref{2a}, each term on the right hand side can be simulated by the antiferromagnetic interaction and an appropriate choice of $t'$, that is, $e^{i H_{o}t}$ using Eq.~\eqref{t'}.

Next, using Trotterization, we can then approximate the time evolution under the general Hamiltonian Eq.~\eqref{EQ_Hf} in terms of the time evolutions under $H_{o}$ and $H_e$ as
\begin{align}
\label{uft}
U_{f,n}(t)&\equiv\left(e^{-i\frac{H_{o}}{2}\tau'}e^{-iH_{e}\tau'}e^{-i\frac{H_{o}}{2}\tau'}\right)^N\ ,
\end{align}
where $\tau'=\frac{\tilde{J}}{J} \left( \frac{2 \pi}{\tilde{J} |\Delta \epsilon|} - \frac{t}{N}\right)$, as per Eq.~\eqref{t'}. Note that we here assume that the $J_{i,i+1}$'s take the same value $J = \tilde{J}$ when turned on. One can adjust individual gate timings to correct for this if they are of different amplitudes. With these choices, we have $U_{f,n} \approx \exp(-i H_{f,n} t)$ up to a small Trotter error. Therefore, we have managed to simulate Eq.~\eqref{EQ_Hf} for $n$ spins using the underlying antiferromagnetic interaction.

In order to demonstrate and verify our technique of simulating ferromagnetic interaction, we further propose two protocols in the following sections, namely the Loschmidt echo protocol and the perfect state transfer protocol.
\section{Loschmidt echo}
\subsection{Loschmidt echo}
Here we are going to use our simulation technique to perform a Loschmidt echo~\cite{peres84,jala01}. With an initial state $\ket{\psi(0)}$ and some Hamiltonians $H_1$ and $H_2$, a Loschmidt echo process is defined as
\begin{align}
\ket{\psi(2t)}=e^{-iH_2t}e^{-iH_1t}\ket{\psi(0)},
\end{align}
where the time evolution operations of $H_1$ and $H_2$ are successively applied to $\ket{\psi(0)}$ for a same time period $t$.
When $H_1=-H_2=H$, the two processes $e^{-iH_1t}$ and $e^{-iH_2t}$ correspond to forward and backward evolutions under the same Hamiltonian. This time reversal process will result in  a revival of the initial state $\ket{\psi(2t)}=\ket{\psi(0)}$. This is the Loschmidt echo. We notice that the relation between $H_2$ and $H_1$ corresponds well with the systems we are working on: 
\begin{align}
H_{a,n}=-H_{f,n}\label{eH}.
\end{align}
Therefore, if we manage to simulate ferromagnetic and antiferromagnetic interactions that satisfy Eq.~\eqref{eH}, we can obtain a revival of the initial state within the Loschmidt echo protocol.

Our Loschmidt echo protocol is as follows. We first prepared $n$ spins in the initial state
\begin{align}
\ket{\psi(0)}=\ket{s}\ket{000...0},
\end{align}
where $ \ket{s} $ is a singlet for the first two spins and $ \ket{000...0} $ represents spin-ups for the other $n-2$ following spins. Experimentally, this state is easy to prepare and the choice of singlet for the first two spins help us not only confirm ferromagnet but also rule out a classical simulator. We choose to turn off the interaction between the first two spins in the following processes, leaving the first spin as a reference.

Next, let the initial state $ \ket{\psi(0)} $ evolve for exactly the same time  period for time $ t $ under $ H_{f,n} $ and $ H_{a,n} $ successively
\begin{align}
\ket{\psi(t)}&=U_{f,n}(t)\ket{\psi(0)},\\
\ket{\psi(2t)}&=U_{af,n}(t)\ket{\psi{(t)}}\\
&\label{ele}=U_{af,n}(t)U_{f,n}(t)\ket{\psi(0)},
\end{align}
with 
\begin{align}
\label{uuu}
U_{af,n}(t)&=\left(e^{-i\frac{H_{o}}{2}\frac{t}{N}}e^{-iH_{e}\frac{t}{N}}e^{-i\frac{H_{o}}{2}\frac{t}{N}}\right)^N,
\end{align}
where we take all $J_{i,i+1},\tilde J_{i,i+1}$ $(1<i<n)$ equal to $J$, and $J_{1,2},\tilde J_{1,2}=0$ during the relevant steps.
Here we note that although such $U_{af,n}(t)$ can in principle be implemented continuously by the simulator, using Trotterization for this unitary exactly cancels out the Trotterization errors introduced during the simulation of the ferromagnetic unitary $U_{f,n}(t)$. We later confirm this observation in our numerical result. Note also that, the interaction between the first two spins is turned off. Therefore, in our Loschmidt echo protocol, $ H_o $ does not contain the first term in Eq.~\eqref{ho}, 
\begin{align}
\label{hto}
H_o=J\sum_{i=2}^{2\lfloor\frac{n}{2}\rfloor-1} \bm{S}_{2i-1}\cdot\bm{S}_{2i},
\end{align}
where $\lfloor\frac{n}{2}\rfloor$ is the floor function of $\frac{n}{2}$.
$H_e$ is still the same as in Eq.~\eqref{he}.

In Eq.~\eqref{ele}, $U_{af}(t)$, the time evolution under the antiferromagnetic interaction, can be directly applied due to our assumption and from the last section we know that we can simulate $U_{f,n}(t)$ by an antiferromagnet. Therefore, the whole Loschmidt echo process in Eq.~\eqref{ele} can be realized by experiment.

\begin{figure}[t]
	\includegraphics[width=0.4\textwidth]{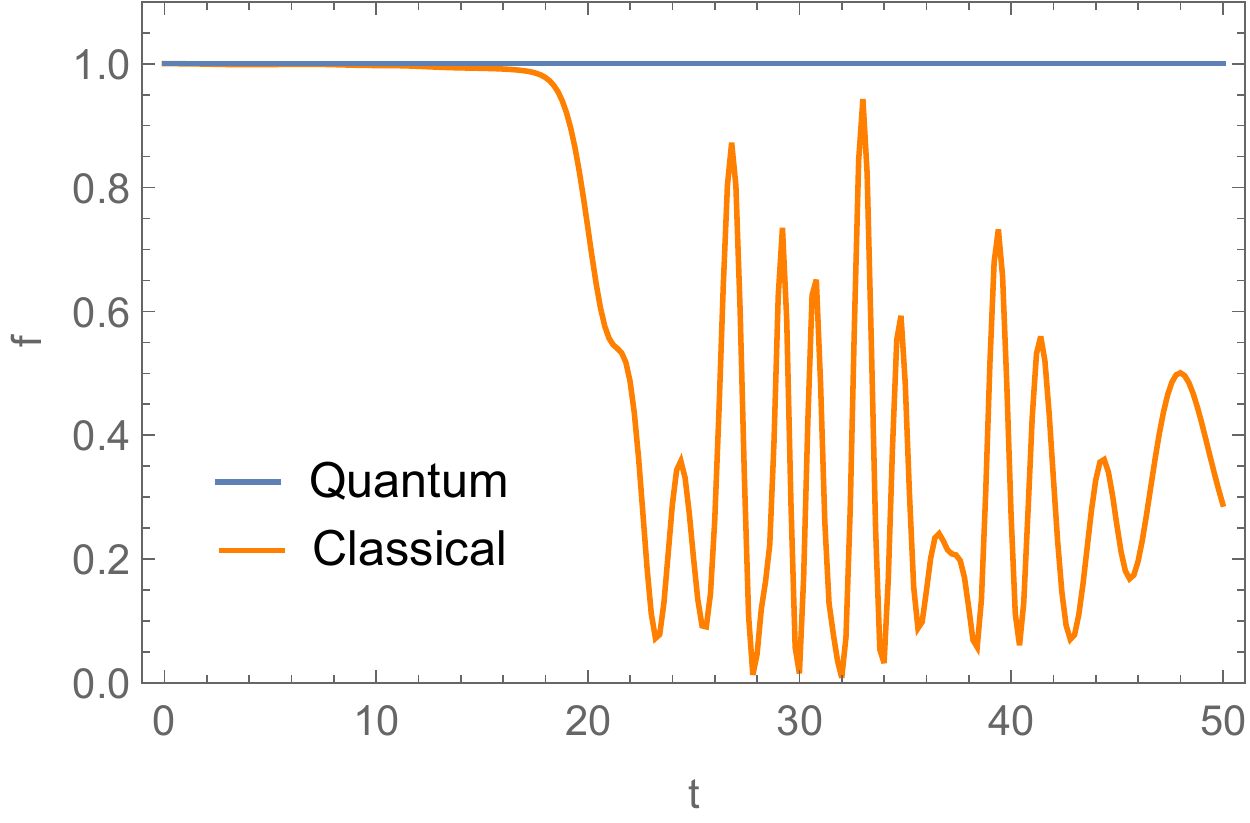}
	\caption{The echo fidelity of quantum (blue) and classical (orange) simulations for 10-spin chain, as a function of evolution time $ t $ ($J$ was taken to be 1).}
	\label{echo}
\end{figure}

To quantify the success of a Loschmidt echo evolution in our protocol, we define the (effective) fidelity of an echo process to be the projection of the first two spins on the singlet state at the final time $2t$:
\begin{align}
\label{fec}
f_{ec}(t)\equiv\bra{\psi(2t)}P_{ec}\ket{\psi(2t)},
\end{align}
where $P_{ec}=\ket{s}\bra{s}\otimes\mathbbm{1}$ is the projection operator. 
If Eq.~\eqref{eH} is satisfied by our simulation method, the revival of the singlet state will be achieved with $f_{ec}(t) = 1$ for all evolution time $t$.

\subsection{Classical model}

As we have mentioned, the choice of initializing the first two spins in a singlet state allows us to confirm certain quantum behaviors of the simulator. Indeed, since the singlet is entangled, revival is not guaranteed in a classical mean field approximation where the two-body interactions in Eq.~\eqref{EQ_Hf} are approximated by local Hamiltonians on individual spins, i.e.
\begin{align}
	H_i(t) = \bm h_i(t)\cdot\bm S_{i}, 
\end{align}
with $\bm{h}_i(t)$ being the mean field experienced by the $i$th spin. 
For the Hamiltonian in Eq.~\eqref{EQ_Hf}, the mean fields are:
\begin{align}
\label{hm1}
	\bm h_2(t) &=J_{2,3}\langle\bm S_3(t)\rangle, \\
    \label{hm3}
    \bm h_i(t) &=J_{i-1,i}\langle\bm S_{i-1}(t)\rangle+J_{i,i+1}\langle\bm S_{i+1}(t)\rangle, \\
    \label{hm2}
    \bm h_n(t)&=J_{n-1,n}\langle\bm S_{n-1}(t)\rangle,
\end{align}
where Eq.~\eqref{hm3} is for $i\in[3,n-1]$.



Since the initial state is a product state of a singlet and $n-2$ spin-ups, the Hamiltonians in Eq.~\eqref{hm1}-\eqref{hm2} result in a system of $n-1$ time-dependent coupled differential equations. 
One of the equations is
\begin{align}
	i\frac{\partial}{\partial t}\ket{\psi_{12}(t)} =  H_{2}(t)\ket{\psi_{12}(t)},
\end{align}
where $\ket{\psi_{12}(t)}$ is the state of the first two spins at time $t$ and $\ket{\psi_{12}(0)}=\ket{s}$. 
The other $n-2$ equations are
\begin{align}
	i\frac{\partial}{\partial t}\ket{\psi_{i}(t)} =  H_{i}(t)\ket{\psi_{i}(t)},
\end{align}
with $\ket{\psi_{i}(t)}$ being the state of the $i$th qubit at time $t$ and $\ket{\psi_{i}(0)}=\ket{0}$.
We numerically solve the coupled equations using Runge-Kutta method to find the state of the $n$ spins at time $t$. 
We then measure the same projection of the first two spins and obtain the fidelity as in Eq.~\eqref{fec}.
\begin{figure}[t]
	\includegraphics[width=0.45\textwidth]{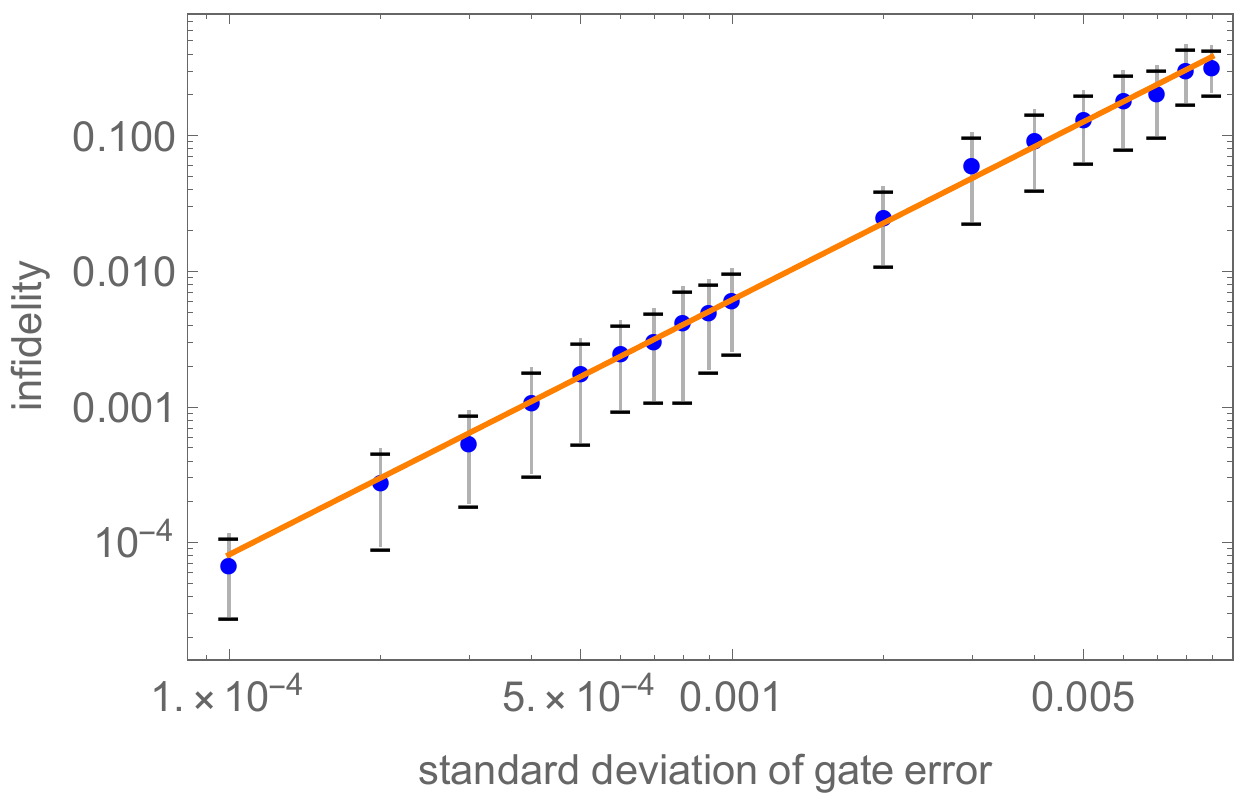}
        \caption{Loschmidt echo infidelity $I_{ec}$ at $n = 25, t = \frac{\pi}{2}$ as a function of the standard deviation $v$ of gate error. The mean values (dots) and standard deviation of $I_{ec}$ are obtained by repeating the simulation 100 times for each value of $v$. Using the fit function 
        $I_{ec}=\exp(a) v^b$, we can find the slope $ b(n) $ for each $n$.}
        \label{re1}
\end{figure}  
      
\begin{figure}[t]        
 	\includegraphics[width=0.45\textwidth]{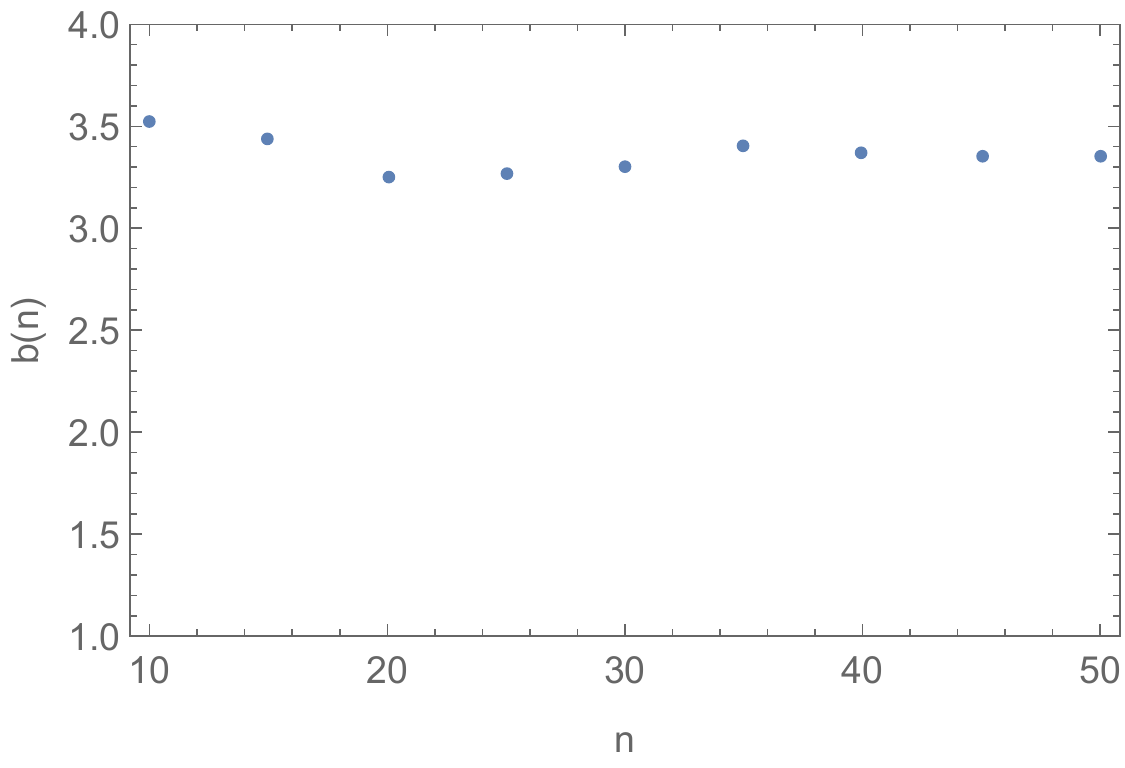}
		
        \caption{The slope $ b(n) $ from the fit function in Fig.~\ref{re1} as a function of $n$.}
        \label{re2}
\end{figure}
\subsection{Numerical result}
The numerical results for both the quantum and classical cases within the Loschmidt echo protocol are shown in Fig.~\ref{echo}. 
The fidelity in quantum condition maintains to be 1 for all evolution time t.
This indicates that our simulation process satisfies Eq.~\eqref{eH} and our simulation of ferromagnet with antiferromagnet is a success. In comparison, when we apply the mean-field approximation as described in last section, there is a deviation from the perfect fidelity. This obvious difference between how the fidelities vary as evolution time grows provides a possible verification of a claimed quantum simulator.
We also note the reason why there is no Trotterization error in the numerical fidelity of the quantum model is because Trotterization is applied to simulate both the ferromagnetic and the antiferromagnetic time evolution unitaries as we have discussed earlier. 
If we instead apply Trotterization on only the ferromagnetic unitary $U_f(t)$ and simulate the antiferromagnetic one continuously without Trotterization, the fidelity is expected to deviate from 1 as time grows due to Trotterization error.
This deviation therefore serves as a possible measure of Trotterization error in the quantum simulator.

\subsection{Robustness}

In experiment, the gate error will be  inevitably involved and effect the simulation. Here we consider the gate error as an additional term of exchange energy for $ J_{i,i+1} $ (Eq.~\eqref{EQ_Ha}), expressed by $ J_{i,i+1} (1+\eta_i) $ ($\eta_i\ll1$), where $ \eta_i $ is a random error sampled from a normal distribution 
\begin{align}
\label{nn}
p(\eta_i)=\frac{e^{-\frac{x^2}{2v^2}}}{\sqrt{2\pi v^2}}, 
\end{align}
of a standard deviation $ v $.  For each step of the time evolution in this series, a new $\eta_i$ is sampled from the distribution above. We ran the numerical experiments for 100 times for different $n$. Recall that in our simulation technique, the time evolution $U_{f,n}(t)$ (Eq.~\eqref{uft}) is approximated by a series of time evolutions under two-spin ferromagnetic Hamiltonians which are relatively long (nearly a $2\pi$ phase evolution) for each Trotter step. In contrast, the return under $U_{af,n}$ uses short steps. This discrepancy will lead to the potential for large errors under small variations of $J_{i,i+1}$.

Here we define infidelity $I_{ec}=1-f_{ec}$ as a measure of imperfect revival. For each $v,n$, we repeat the numerical simulation 100 times to get an averaged infidelity. In Fig.~\ref{re1}, we plot this averaged infidelity for several choice of $v$ at fixed $n=25$. The plots shows that infidelity grows only polynomially with $v$ for a fixed $n$, i.e. $ I_{ec}\propto\sigma^b $ for some order $b$. By taking log-log plot and finding the linear fit of it as $\log (I_{ec})=a+b(n) \log(v) $, we obtained the slope $b(n)$ of Fig.~\ref{re1} and we further define order $b(n)$ as the robustness of the specific systems. In Fig.~\ref{re2}, the slope $b(n)$ is plotted as a function of $n$. We see that $b$ is largely independent of $n$, and thus we can conclude that this protocol is 'robust' in the sense that the fidelity does not decrease exponentially with increasing numbers of spins. Regarding the source of this robustness, we note that our protocol does not necessarily send the spin information through arbitrary distances in an infinite chain, possibly due Anderson localization in our one dimensional system.


\section{Perfect state transfer}
\subsection{State transfer}
\begin{figure}[t]
	\centering
	\includegraphics[scale=0.35]{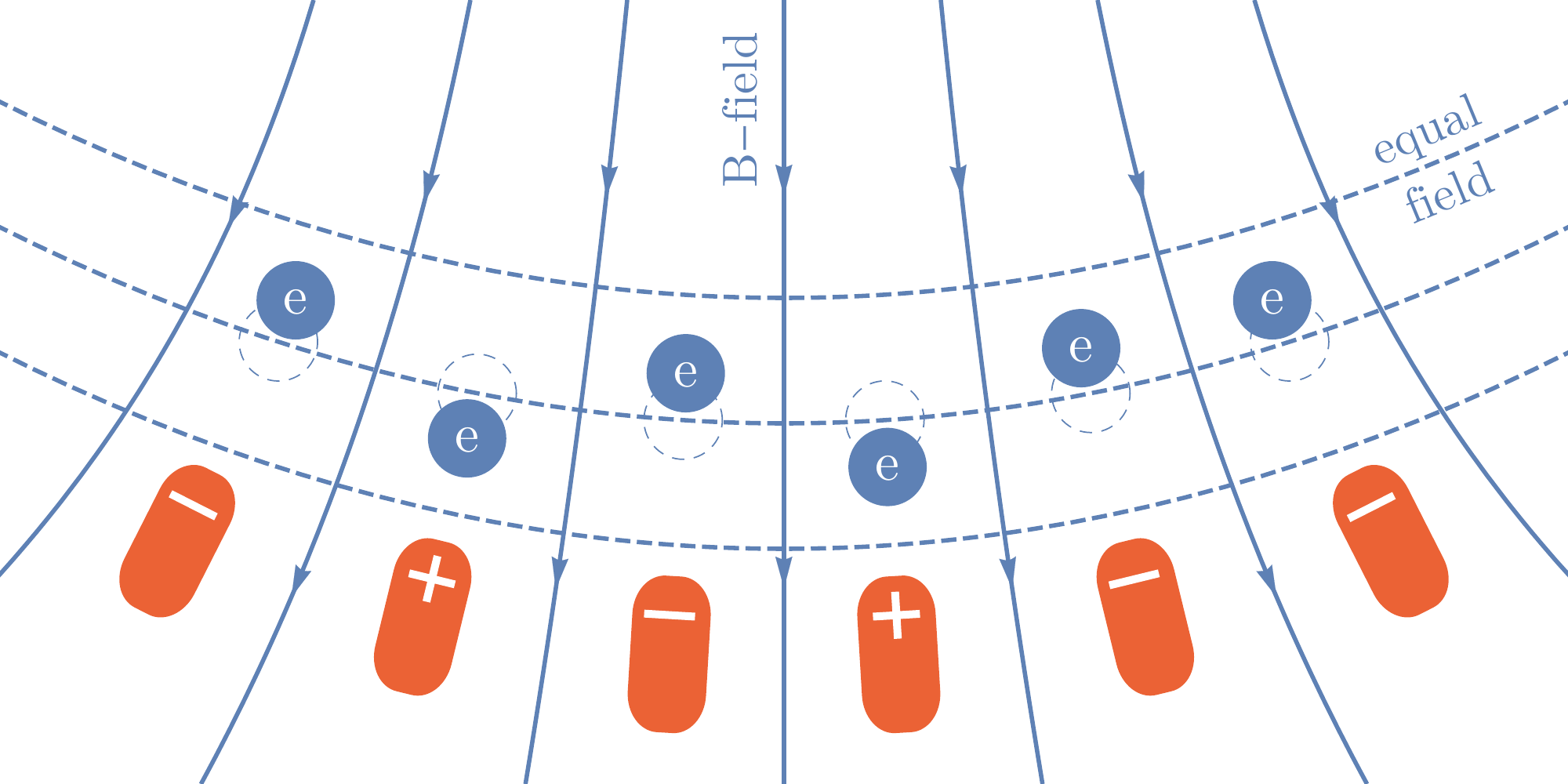}
	\caption{This is the illustration of the architecture of magnetic field gradient in experiment. Spins are aligned on a equal field line of the magnetic field of a magnetic dipole and each spin is manipulated by separated electrodes so that they can move in the field and finally line up in a gradient field.}
	\label{B}
\end{figure}

The Loschmidt echo protocol provides us with a verification of the existence of ferromagnetic interaction in our simulation.  However, since Loschmidt echo only gives the measure result of the first two spins, there is no guarantee that the information is transferring throughout the whole spin chain, especially from one end to the other. A perfect state transfer from one end to the other can be achieved under the Hamiltonian \cite{StateTransfer}
\begin{align}
\label{ht}
H_{\text{tr}}=-2 \sum_{i=1}^{n-1}J_{i,i+1}\bm{S}_i\cdot\bm{S}_{i+1} + \sum_{i=1}^{n} B_i\sigma^z.
\end{align}
Compared to the Heisenberg model for ferromagnetic interaction we used, the differences included are a non-uniform exchange interaction between the $ i $th and the $(i+1)$th spins
\begin{align}
\label{jj}
J_{i,i+1}=\sqrt{i (n-i)}, 
\end{align}
and a nonuniform magnetic field
\begin{align}
B_i=\frac{1}{2}(J_{i,i+1}+J_{i-1,i}),
\end{align}
on the $i$th spin.
Such a nonuniform magnetic field can be engineered using the architecture illustrated in Fig.~\ref{B}. The spins can be realized as electrons in quantum dots placed in a magnetic field gradient. The magnetic field strength on each spin can be adjusted using an electrode which may pull or push the electron to a different magnetic field strength. With this extra magnetic field term in the state transfer Hamiltonian Eq.~\eqref{ht}, we split it into three Trotter elements, i.e. $H_o$, $H_e$ and the magnetic field term $H_B=\sum_{i=1}^{n} B_i\sigma^z$.
Under the Hamiltonian in Eq.~\eqref{ht}, the initial state $ \ket{\psi(0)}=\ket{s}\ket{000...0} $ would evolve to $\ket{ \psi(\frac{\pi}{2})}=\ket{000...0}\ket{s} $ after a time $t=\frac{\pi}{2}$.  
\begin{figure}[t]
	\centering
	\includegraphics[scale=0.5]{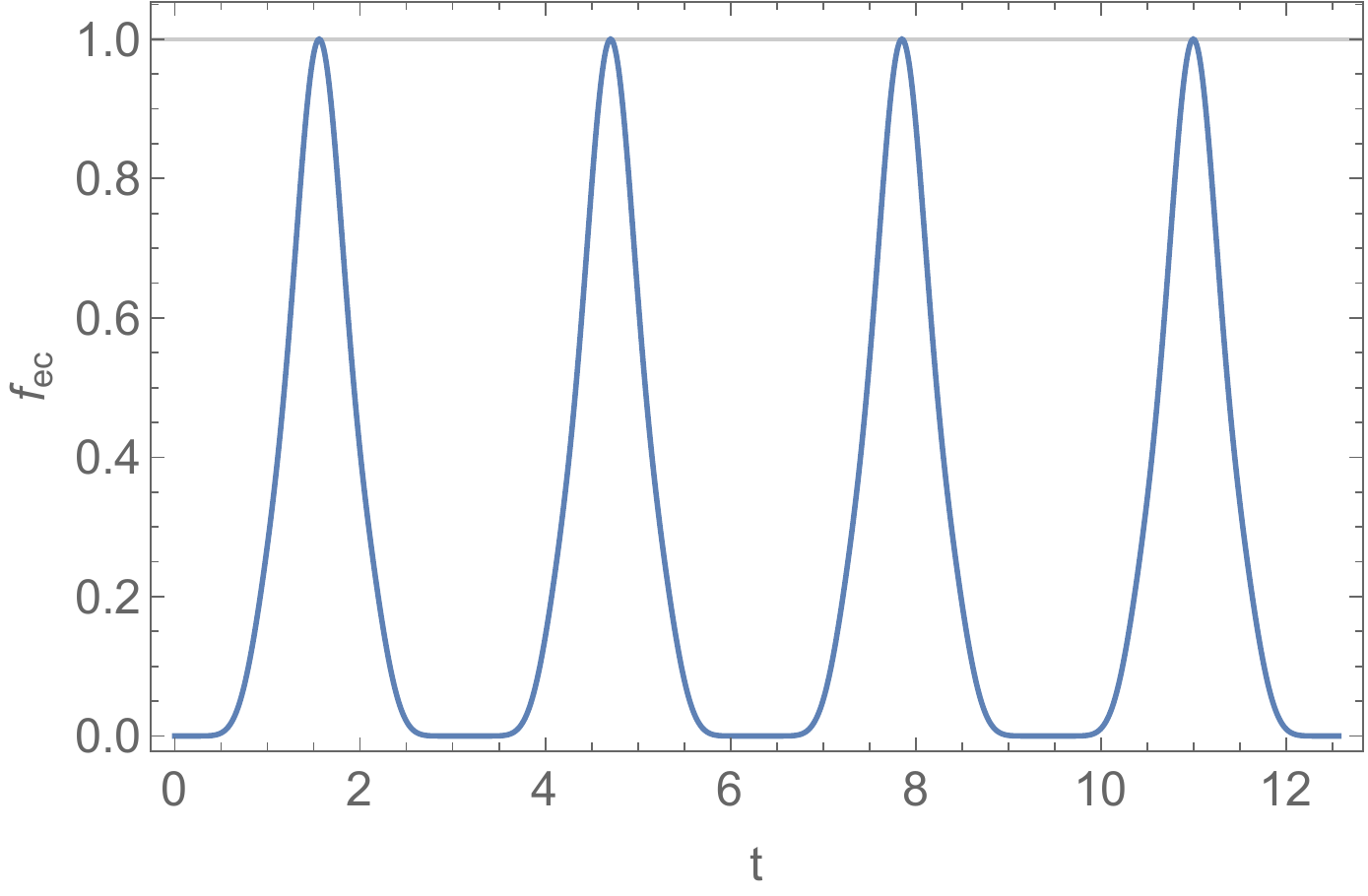}
	\caption{The figure shows the fidelity in our state transfer protocol as a function of the evolution time $ t $. The state is perfectly transfered through the chain at $ t=\frac{\pi}{2} $.}
	\label{st}
\end{figure}

\begin{figure}[t]

		\includegraphics[width=0.45\textwidth]{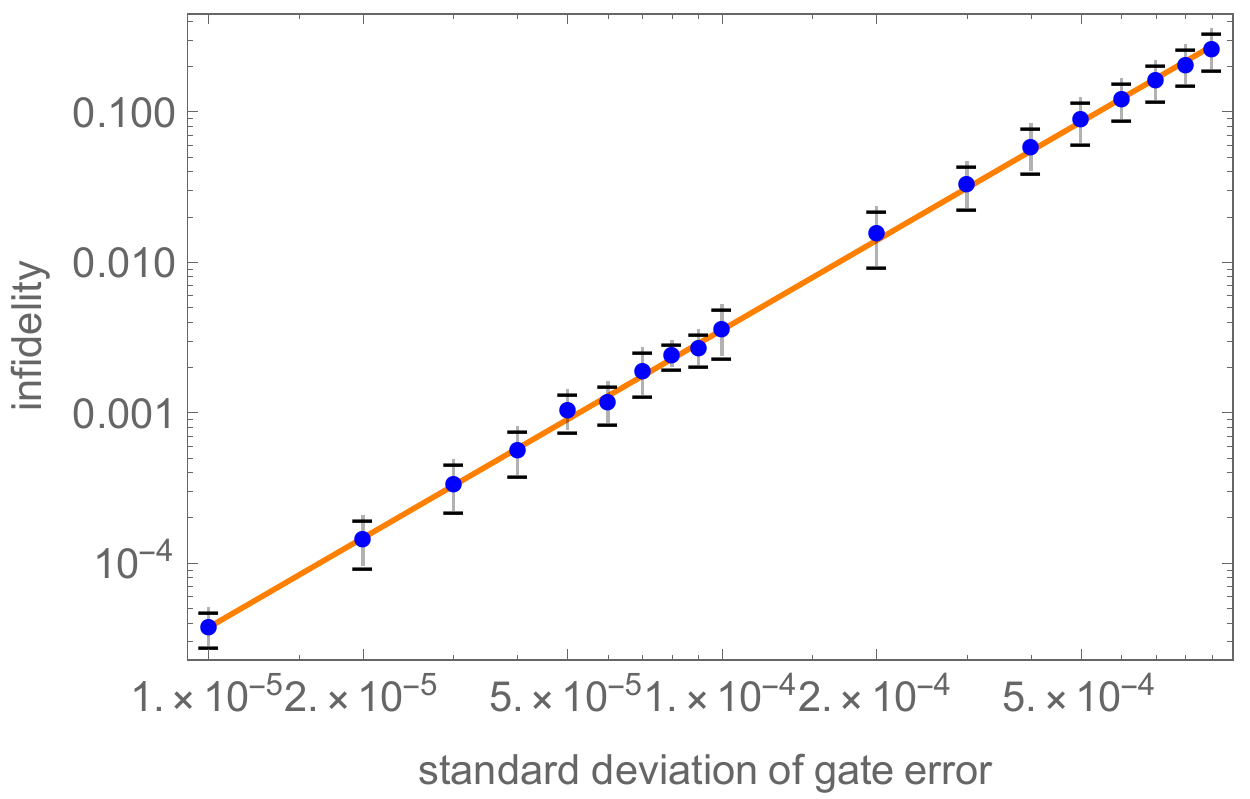}
	        \caption{Perfect state transfer infidelity $I_{tr}$ at $n = 25, t = \frac{\pi}{2}$ as a function of the standard deviation $v$ of gate error. The mean values (dots) and standard deviations of $I_{tr}$ are obtained by repeating the simulation 100 times for each value of $v$. Using the fit function $I_{tr}=\exp(a)v^b $, we can find the slope $ b(n) $ for each $n$.}
            \label{rs1}
\end{figure}
\begin{figure}

\includegraphics[width=0.43\textwidth]{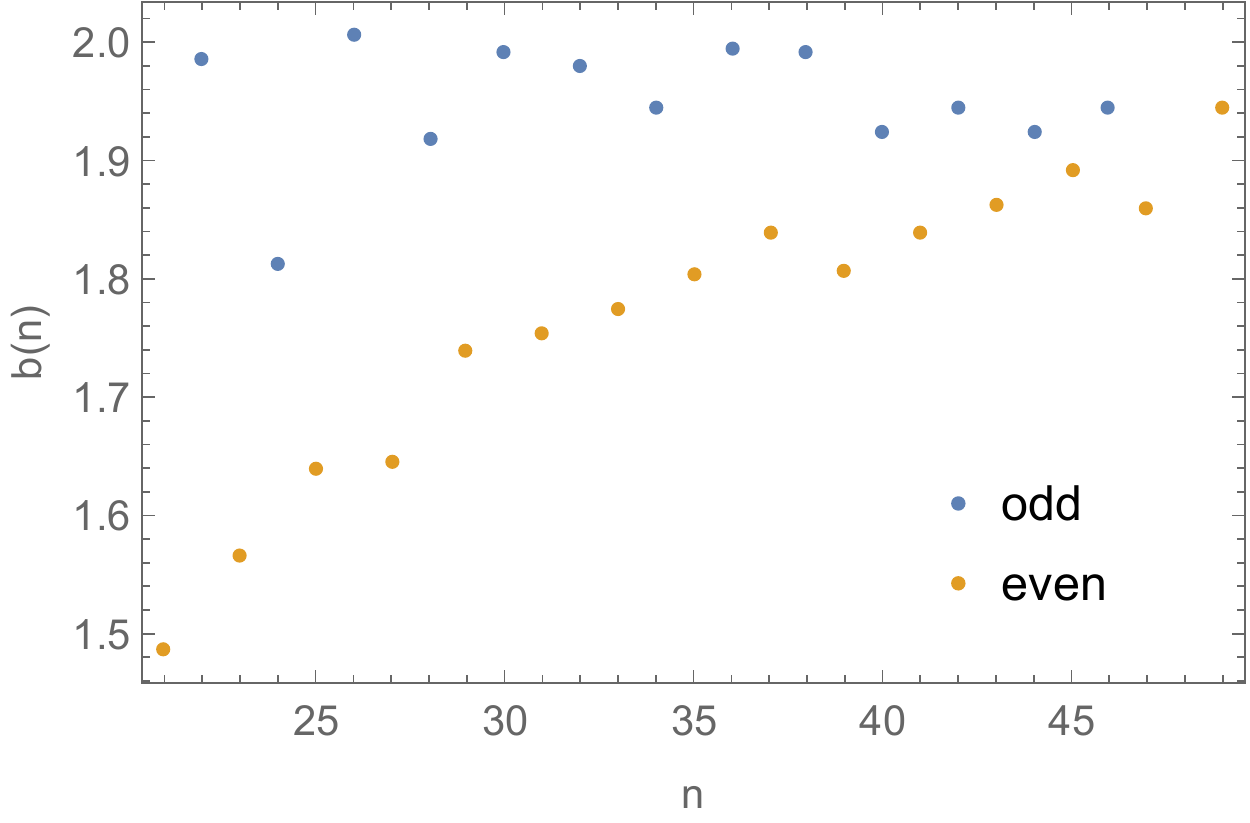}
        \caption{The slope $ b(n) $ from the fit function in Fig.~\ref{rs1} as a function of $n$. $b(n)$ for even $n$ (orange) and odd $n$ (blue) show different dependence of $n$.}
        \label{rs2}
\end{figure}
		
Note that when we apply our method of simulating ferromagnet with antiferromagnet, the different $J_i$ between spins will cause different simulation time interval $t'$ under the antiferromagnetic interactions. We therefore simulate each pair separately. These simulations can be done in parallel and hence do not change our Trotterization choice. 
After evolving for the time $ t=\frac{\pi}{2} $ under $ H_{\text{tr}} $, we are expecting a final state $\ket{ \psi(\frac{\pi}{2})}=\ket{000...0}\ket{s} $. We measure the projection of the last two spins onto $ \ket{s} $ using
\begin{align}
P_{tr}=\mathbbm{1}\otimes\ket{s}\bra{s}.
\end{align}
We further define the fidelity $f_{tr}$ of this protocol to be
\begin{align}
f_{tr}\equiv\bra{\psi\left(\frac{\pi}{2}\right)}P_{tr}\ket{\psi\left(\frac{\pi}{2}\right)}.
\end{align}
The numerical result in Fig.~\ref{st} shows that the state is transferred throughout the whole chain perfectly at $ t=\frac{\pi}{2} $. This provides us with the confidence to say that the interaction we introduced to the simulation is the nearest-neighbor two-body interactions we mean to simulate.

\subsection{Robustness}
Similar to the robustness for Loschmidt echo discussed above, we add the gate error term to Eq.~\eqref{jj} and get $ J_{i,i+1}(1+\eta_i) $($\eta_i\ll1$), where $ \eta_i $ is a random error sampled from the normal distribution in Eq.~\eqref{nn}. The time evolution under $H_{tr}$ (Eq.~\eqref{ht}) is approximated by a series of time evolutions under two-spin ferromagnetic Hamiltonians. For each time evolution in this series, a new $\eta_i$ is sampled from the distribution above. 

We define infidelity $I_{tr}=1-f_{tr}$ as the deviation of the fidelity from the perfect value 1. For each $v$ and $n$, we repeated the numerical simulation 100 times to get an averaged infidelity. Fig.~\ref{rs1} describes how infidelity changes with the standard deviation $v$ at $n=25$. The linear fit $\log I_{tr}=a+b \log v$ of a log-log plot gives the slope $b(n)$ which tells us how fast the infidelity grows with the error strength. In Fig.~\ref{rs2}, we plot how $b(n)$ changes with $n$. It indicates that, the robustness in state transfer protocol takes on different pattern in response to $n$. For odd $n$, the system appears to be robust, i.e., $b(n)$ does not have exponentially bad performance as a function of $n$. However, for even $n$ this is no longer the case. We attribute this to the role a single, bad link plays in the even case right in the center of the chain --- it could be this worst case scenario that dominates the success or failure of the transfer protocol. In contrast for odd $n$, two links are equally strong in the center, leading to multiple failure pathways and (possibly) our source of exponentially decreasing fidelity.

\section{Outlook}
In this paper we consider two tests of a quantum simulator with Heisenberg interactions and the ability to prepare and measure singlet states of spins. Starting with a technique to simulate ferromagnetic interactions using antiferromagnetic interactions, our protocols are surprisingly robust to parametric errors in operation of the simulator.
However, properties of the system in the middle of these protocols have not been investigated, nor have scenarios in which depolarizing noise or state preparation error play a key role. We also do not yet have a way to estimate how much and how fast entanglement entropy grows in the system as a function of time. Answering these questions are intriguing future directions of research.
On the other hand, our state transfer protocol has demonstrated the ability to transport quantum entanglement between subsystems. 
We suspect that a more complicated quantum computation tasks can be also implemented, which might be able to lower bound the computational power of our proposed simulator.

\acknowledgements
We thank X.~Wu and A.~M.~Childs for helpful discussions. This research was supported in part by the NSF funded Physics Frontier Center at the Joint Quantum Institute.

\bibliographystyle{apsrev4-1}
\bibliography{References}

\begin{thebibliography}{34}%
\makeatletter
\providecommand \@ifxundefined [1]{%
 \@ifx{#1\undefined}
}%
\providecommand \@ifnum [1]{%
 \ifnum #1\expandafter \@firstoftwo
 \else \expandafter \@secondoftwo
 \fi
}%
\providecommand \@ifx [1]{%
 \ifx #1\expandafter \@firstoftwo
 \else \expandafter \@secondoftwo
 \fi
}%
\providecommand \natexlab [1]{#1}%
\providecommand \enquote  [1]{``#1''}%
\providecommand \bibnamefont  [1]{#1}%
\providecommand \bibfnamefont [1]{#1}%
\providecommand \citenamefont [1]{#1}%
\providecommand \href@noop [0]{\@secondoftwo}%
\providecommand \href [0]{\begingroup \@sanitize@url \@href}%
\providecommand \@href[1]{\@@startlink{#1}\@@href}%
\providecommand \@@href[1]{\endgroup#1\@@endlink}%
\providecommand \@sanitize@url [0]{\catcode `\\12\catcode `\$12\catcode
  `\&12\catcode `\#12\catcode `\^12\catcode `\_12\catcode `\%12\relax}%
\providecommand \@@startlink[1]{}%
\providecommand \@@endlink[0]{}%
\providecommand \url  [0]{\begingroup\@sanitize@url \@url }%
\providecommand \@url [1]{\endgroup\@href {#1}{\urlprefix }}%
\providecommand \urlprefix  [0]{URL }%
\providecommand \Eprint [0]{\href }%
\providecommand \doibase [0]{http://dx.doi.org/}%
\providecommand \selectlanguage [0]{\@gobble}%
\providecommand \bibinfo  [0]{\@secondoftwo}%
\providecommand \bibfield  [0]{\@secondoftwo}%
\providecommand \translation [1]{[#1]}%
\providecommand \BibitemOpen [0]{}%
\providecommand \bibitemStop [0]{}%
\providecommand \bibitemNoStop [0]{.\EOS\space}%
\providecommand \EOS [0]{\spacefactor3000\relax}%
\providecommand \BibitemShut  [1]{\csname bibitem#1\endcsname}%
\let\auto@bib@innerbib\@empty
\bibitem [{\citenamefont {Feynman}(1982)}]{feynman82}%
  \BibitemOpen
  \bibfield  {author} {\bibinfo {author} {\bibfnamefont {R.~P.}\ \bibnamefont
  {Feynman}},\ }\href@noop {} {\bibfield  {journal} {\bibinfo  {journal}
  {International Journal of Theoretical Physics}\ }\textbf {\bibinfo {volume}
  {21}},\ \bibinfo {pages} {467} (\bibinfo {year} {1982})}\BibitemShut
  {NoStop}%
\bibitem [{\citenamefont {Houck}\ \emph {et~al.}(2012)\citenamefont {Houck},
  \citenamefont {T{\"u}reci},\ and\ \citenamefont {Koch}}]{houck12}%
  \BibitemOpen
  \bibfield  {author} {\bibinfo {author} {\bibfnamefont {A.~A.}\ \bibnamefont
  {Houck}}, \bibinfo {author} {\bibfnamefont {H.~E.}\ \bibnamefont
  {T{\"u}reci}}, \ and\ \bibinfo {author} {\bibfnamefont {J.}~\bibnamefont
  {Koch}},\ }\href@noop {} {\bibfield  {journal} {\bibinfo  {journal} {Nature
  Physics}\ }\textbf {\bibinfo {volume} {8}},\ \bibinfo {pages} {292} (\bibinfo
  {year} {2012})}\BibitemShut {NoStop}%
\bibitem [{\citenamefont {Ma}\ \emph {et~al.}(2014)\citenamefont {Ma},
  \citenamefont {Daki{\'c}}, \citenamefont {Kropatschek}, \citenamefont
  {Naylor}, \citenamefont {Chan}, \citenamefont {Gong}, \citenamefont {Duan},
  \citenamefont {Zeilinger},\ and\ \citenamefont {Walther}}]{ma14}%
  \BibitemOpen
  \bibfield  {author} {\bibinfo {author} {\bibfnamefont {X.}~\bibnamefont
  {Ma}}, \bibinfo {author} {\bibfnamefont {B.}~\bibnamefont {Daki{\'c}}},
  \bibinfo {author} {\bibfnamefont {S.}~\bibnamefont {Kropatschek}}, \bibinfo
  {author} {\bibfnamefont {W.}~\bibnamefont {Naylor}}, \bibinfo {author}
  {\bibfnamefont {Y.}~\bibnamefont {Chan}}, \bibinfo {author} {\bibfnamefont
  {Z.}~\bibnamefont {Gong}}, \bibinfo {author} {\bibfnamefont {L.}~\bibnamefont
  {Duan}}, \bibinfo {author} {\bibfnamefont {A.}~\bibnamefont {Zeilinger}}, \
  and\ \bibinfo {author} {\bibfnamefont {P.}~\bibnamefont {Walther}},\
  }\href@noop {} {\bibfield  {journal} {\bibinfo  {journal} {Scientific
  Reports}\ }\textbf {\bibinfo {volume} {4}},\ \bibinfo {pages} {3583}
  (\bibinfo {year} {2014})}\BibitemShut {NoStop}%
\bibitem [{\citenamefont {O'Malley}\ \emph {et~al.}(2016)\citenamefont
  {O'Malley}, \citenamefont {Babbush}, \citenamefont {Kivlichan}, \citenamefont
  {Romero}, \citenamefont {McClean}, \citenamefont {Barends}, \citenamefont
  {Kelly}, \citenamefont {Roushan}, \citenamefont {Tranter}, \citenamefont
  {Ding}, \citenamefont {Campbell}, \citenamefont {Chen}, \citenamefont {Chen},
  \citenamefont {Chiaro}, \citenamefont {Dunsworth}, \citenamefont {Fowler},
  \citenamefont {Jeffrey}, \citenamefont {Lucero}, \citenamefont {Megrant},
  \citenamefont {Mutus}, \citenamefont {Neeley}, \citenamefont {Neill},
  \citenamefont {Quintana}, \citenamefont {Sank}, \citenamefont {Vainsencher},
  \citenamefont {Wenner}, \citenamefont {White}, \citenamefont {Coveney},
  \citenamefont {Love}, \citenamefont {Neven}, \citenamefont {Aspuru-Guzik},\
  and\ \citenamefont {Martinis}}]{omalley16}%
  \BibitemOpen
  \bibfield  {author} {\bibinfo {author} {\bibfnamefont {P.~J.~J.}\
  \bibnamefont {O'Malley}}, \bibinfo {author} {\bibfnamefont {R.}~\bibnamefont
  {Babbush}}, \bibinfo {author} {\bibfnamefont {I.~D.}\ \bibnamefont
  {Kivlichan}}, \bibinfo {author} {\bibfnamefont {J.}~\bibnamefont {Romero}},
  \bibinfo {author} {\bibfnamefont {J.~R.}\ \bibnamefont {McClean}}, \bibinfo
  {author} {\bibfnamefont {R.}~\bibnamefont {Barends}}, \bibinfo {author}
  {\bibfnamefont {J.}~\bibnamefont {Kelly}}, \bibinfo {author} {\bibfnamefont
  {P.}~\bibnamefont {Roushan}}, \bibinfo {author} {\bibfnamefont
  {A.}~\bibnamefont {Tranter}}, \bibinfo {author} {\bibfnamefont
  {N.}~\bibnamefont {Ding}}, \bibinfo {author} {\bibfnamefont {B.}~\bibnamefont
  {Campbell}}, \bibinfo {author} {\bibfnamefont {Y.}~\bibnamefont {Chen}},
  \bibinfo {author} {\bibfnamefont {Z.}~\bibnamefont {Chen}}, \bibinfo {author}
  {\bibfnamefont {B.}~\bibnamefont {Chiaro}}, \bibinfo {author} {\bibfnamefont
  {A.}~\bibnamefont {Dunsworth}}, \bibinfo {author} {\bibfnamefont {A.~G.}\
  \bibnamefont {Fowler}}, \bibinfo {author} {\bibfnamefont {E.}~\bibnamefont
  {Jeffrey}}, \bibinfo {author} {\bibfnamefont {E.}~\bibnamefont {Lucero}},
  \bibinfo {author} {\bibfnamefont {A.}~\bibnamefont {Megrant}}, \bibinfo
  {author} {\bibfnamefont {J.~Y.}\ \bibnamefont {Mutus}}, \bibinfo {author}
  {\bibfnamefont {M.}~\bibnamefont {Neeley}}, \bibinfo {author} {\bibfnamefont
  {C.}~\bibnamefont {Neill}}, \bibinfo {author} {\bibfnamefont
  {C.}~\bibnamefont {Quintana}}, \bibinfo {author} {\bibfnamefont
  {D.}~\bibnamefont {Sank}}, \bibinfo {author} {\bibfnamefont {A.}~\bibnamefont
  {Vainsencher}}, \bibinfo {author} {\bibfnamefont {J.}~\bibnamefont {Wenner}},
  \bibinfo {author} {\bibfnamefont {T.~C.}\ \bibnamefont {White}}, \bibinfo
  {author} {\bibfnamefont {P.~V.}\ \bibnamefont {Coveney}}, \bibinfo {author}
  {\bibfnamefont {P.~J.}\ \bibnamefont {Love}}, \bibinfo {author}
  {\bibfnamefont {H.}~\bibnamefont {Neven}}, \bibinfo {author} {\bibfnamefont
  {A.}~\bibnamefont {Aspuru-Guzik}}, \ and\ \bibinfo {author} {\bibfnamefont
  {J.~M.}\ \bibnamefont {Martinis}},\ }\href {\doibase
  10.1103/PhysRevX.6.031007} {\bibfield  {journal} {\bibinfo  {journal} {Phys.
  Rev. X}\ }\textbf {\bibinfo {volume} {6}},\ \bibinfo {pages} {031007}
  (\bibinfo {year} {2016})}\BibitemShut {NoStop}%
\bibitem [{\citenamefont {Hensgens}\ \emph {et~al.}(2017)\citenamefont
  {Hensgens}, \citenamefont {Fujita}, \citenamefont {Janssen}, \citenamefont
  {Li}, \citenamefont {Van~Diepen}, \citenamefont {Reichl}, \citenamefont
  {Wegscheider}, \citenamefont {Das~Sarma},\ and\ \citenamefont
  {Vandersypen}}]{hensgens17}%
  \BibitemOpen
  \bibfield  {author} {\bibinfo {author} {\bibfnamefont {T.}~\bibnamefont
  {Hensgens}}, \bibinfo {author} {\bibfnamefont {T.}~\bibnamefont {Fujita}},
  \bibinfo {author} {\bibfnamefont {L.}~\bibnamefont {Janssen}}, \bibinfo
  {author} {\bibfnamefont {X.}~\bibnamefont {Li}}, \bibinfo {author}
  {\bibfnamefont {C.~J.}\ \bibnamefont {Van~Diepen}}, \bibinfo {author}
  {\bibfnamefont {C.}~\bibnamefont {Reichl}}, \bibinfo {author} {\bibfnamefont
  {W.}~\bibnamefont {Wegscheider}}, \bibinfo {author} {\bibfnamefont
  {S.}~\bibnamefont {Das~Sarma}}, \ and\ \bibinfo {author} {\bibfnamefont
  {L.~M.~K.}\ \bibnamefont {Vandersypen}},\ }\href
  {http://dx.doi.org/10.1038/nature23022} {\bibfield  {journal} {\bibinfo
  {journal} {Nature}\ }\textbf {\bibinfo {volume} {548}},\ \bibinfo {pages}
  {70} (\bibinfo {year} {2017})}\BibitemShut {NoStop}%
\bibitem [{\citenamefont {Loredo}\ \emph {et~al.}(2016)\citenamefont {Loredo},
  \citenamefont {Almeida}, \citenamefont {Di~Candia}, \citenamefont
  {Pedernales}, \citenamefont {Casanova}, \citenamefont {Solano},\ and\
  \citenamefont {White}}]{loredo16}%
  \BibitemOpen
  \bibfield  {author} {\bibinfo {author} {\bibfnamefont {J.~C.}\ \bibnamefont
  {Loredo}}, \bibinfo {author} {\bibfnamefont {M.~P.}\ \bibnamefont {Almeida}},
  \bibinfo {author} {\bibfnamefont {R.}~\bibnamefont {Di~Candia}}, \bibinfo
  {author} {\bibfnamefont {J.~S.}\ \bibnamefont {Pedernales}}, \bibinfo
  {author} {\bibfnamefont {J.}~\bibnamefont {Casanova}}, \bibinfo {author}
  {\bibfnamefont {E.}~\bibnamefont {Solano}}, \ and\ \bibinfo {author}
  {\bibfnamefont {A.~G.}\ \bibnamefont {White}},\ }\href {\doibase
  10.1103/PhysRevLett.116.070503} {\bibfield  {journal} {\bibinfo  {journal}
  {Phys. Rev. Lett.}\ }\textbf {\bibinfo {volume} {116}},\ \bibinfo {pages}
  {070503} (\bibinfo {year} {2016})}\BibitemShut {NoStop}%
\bibitem [{\citenamefont {Zhang}\ \emph {et~al.}(2017)\citenamefont {Zhang},
  \citenamefont {Pagano}, \citenamefont {Hess}, \citenamefont {Kyprianidis},
  \citenamefont {Becker}, \citenamefont {Kaplan}, \citenamefont {Gorshkov},
  \citenamefont {Gong},\ and\ \citenamefont {Monroe}}]{monroe17}%
  \BibitemOpen
  \bibfield  {author} {\bibinfo {author} {\bibfnamefont {J.}~\bibnamefont
  {Zhang}}, \bibinfo {author} {\bibfnamefont {G.}~\bibnamefont {Pagano}},
  \bibinfo {author} {\bibfnamefont {P.~W.}\ \bibnamefont {Hess}}, \bibinfo
  {author} {\bibfnamefont {A.}~\bibnamefont {Kyprianidis}}, \bibinfo {author}
  {\bibfnamefont {P.}~\bibnamefont {Becker}}, \bibinfo {author} {\bibfnamefont
  {H.}~\bibnamefont {Kaplan}}, \bibinfo {author} {\bibfnamefont {A.~V.}\
  \bibnamefont {Gorshkov}}, \bibinfo {author} {\bibfnamefont {Z.-X.}\
  \bibnamefont {Gong}}, \ and\ \bibinfo {author} {\bibfnamefont
  {C.}~\bibnamefont {Monroe}},\ }\href@noop {} {\bibfield  {journal} {\bibinfo
  {journal} {Nature}\ }\textbf {\bibinfo {volume} {551}},\ \bibinfo {pages}
  {601} (\bibinfo {year} {2017})}\BibitemShut {NoStop}%
\bibitem [{\citenamefont {Bernien}\ \emph {et~al.}(2017)\citenamefont
  {Bernien}, \citenamefont {Schwartz}, \citenamefont {Keesling}, \citenamefont
  {Levine}, \citenamefont {Omran}, \citenamefont {Pichler}, \citenamefont
  {Choi}, \citenamefont {Zibrov}, \citenamefont {Endres}, \citenamefont
  {Greiner} \emph {et~al.}}]{lukin17}%
  \BibitemOpen
  \bibfield  {author} {\bibinfo {author} {\bibfnamefont {H.}~\bibnamefont
  {Bernien}}, \bibinfo {author} {\bibfnamefont {S.}~\bibnamefont {Schwartz}},
  \bibinfo {author} {\bibfnamefont {A.}~\bibnamefont {Keesling}}, \bibinfo
  {author} {\bibfnamefont {H.}~\bibnamefont {Levine}}, \bibinfo {author}
  {\bibfnamefont {A.}~\bibnamefont {Omran}}, \bibinfo {author} {\bibfnamefont
  {H.}~\bibnamefont {Pichler}}, \bibinfo {author} {\bibfnamefont
  {S.}~\bibnamefont {Choi}}, \bibinfo {author} {\bibfnamefont {A.~S.}\
  \bibnamefont {Zibrov}}, \bibinfo {author} {\bibfnamefont {M.}~\bibnamefont
  {Endres}}, \bibinfo {author} {\bibfnamefont {M.}~\bibnamefont {Greiner}},
  \emph {et~al.},\ }\href@noop {} {\bibfield  {journal} {\bibinfo  {journal}
  {Nature}\ }\textbf {\bibinfo {volume} {551}},\ \bibinfo {pages} {579}
  (\bibinfo {year} {2017})}\BibitemShut {NoStop}%
\bibitem [{\citenamefont {Shin}\ \emph {et~al.}(2014)\citenamefont {Shin},
  \citenamefont {Smith}, \citenamefont {Smolin},\ and\ \citenamefont
  {Vazirani}}]{vazirani14}%
  \BibitemOpen
  \bibfield  {author} {\bibinfo {author} {\bibfnamefont {S.~W.}\ \bibnamefont
  {Shin}}, \bibinfo {author} {\bibfnamefont {G.}~\bibnamefont {Smith}},
  \bibinfo {author} {\bibfnamefont {J.~A.}\ \bibnamefont {Smolin}}, \ and\
  \bibinfo {author} {\bibfnamefont {U.}~\bibnamefont {Vazirani}},\ }\href@noop
  {} {\bibfield  {journal} {\bibinfo  {journal} {arXiv preprint
  arXiv:1401.7087}\ } (\bibinfo {year} {2014})}\BibitemShut {NoStop}%
\bibitem [{\citenamefont {Zagoskin}\ \emph {et~al.}(2014)\citenamefont
  {Zagoskin}, \citenamefont {Il'ichev}, \citenamefont {Grajcar}, \citenamefont
  {Betouras},\ and\ \citenamefont {Nori}}]{nori14}%
  \BibitemOpen
  \bibfield  {author} {\bibinfo {author} {\bibfnamefont {A.~M.}\ \bibnamefont
  {Zagoskin}}, \bibinfo {author} {\bibfnamefont {E.}~\bibnamefont {Il'ichev}},
  \bibinfo {author} {\bibfnamefont {M.}~\bibnamefont {Grajcar}}, \bibinfo
  {author} {\bibfnamefont {J.~J.}\ \bibnamefont {Betouras}}, \ and\ \bibinfo
  {author} {\bibfnamefont {F.}~\bibnamefont {Nori}},\ }\href {\doibase
  10.3389/fphy.2014.00033} {\bibfield  {journal} {\bibinfo  {journal}
  {Frontiers in Physics}\ }\textbf {\bibinfo {volume} {2}},\ \bibinfo {pages}
  {33} (\bibinfo {year} {2014})}\BibitemShut {NoStop}%
\bibitem [{\citenamefont {Albash}\ \emph {et~al.}(2015)\citenamefont {Albash},
  \citenamefont {Vinci}, \citenamefont {Mishra}, \citenamefont {Warburton},\
  and\ \citenamefont {Lidar}}]{albash2015}%
  \BibitemOpen
  \bibfield  {author} {\bibinfo {author} {\bibfnamefont {T.}~\bibnamefont
  {Albash}}, \bibinfo {author} {\bibfnamefont {W.}~\bibnamefont {Vinci}},
  \bibinfo {author} {\bibfnamefont {A.}~\bibnamefont {Mishra}}, \bibinfo
  {author} {\bibfnamefont {P.~A.}\ \bibnamefont {Warburton}}, \ and\ \bibinfo
  {author} {\bibfnamefont {D.~A.}\ \bibnamefont {Lidar}},\ }\href@noop {}
  {\bibfield  {journal} {\bibinfo  {journal} {Phys. Rev. A}\ }\textbf {\bibinfo
  {volume} {91}},\ \bibinfo {pages} {042314} (\bibinfo {year}
  {2015})}\BibitemShut {NoStop}%
\bibitem [{\citenamefont {Kafri}\ and\ \citenamefont
  {Taylor}(2015)}]{kafri2015}%
  \BibitemOpen
  \bibfield  {author} {\bibinfo {author} {\bibfnamefont {D.}~\bibnamefont
  {Kafri}}\ and\ \bibinfo {author} {\bibfnamefont {J.}~\bibnamefont {Taylor}},\
  }\href@noop {} {\bibfield  {journal} {\bibinfo  {journal} {arXiv preprint
  arXiv:1504.01187}\ } (\bibinfo {year} {2015})}\BibitemShut {NoStop}%
\bibitem [{\citenamefont {Hangleiter}\ \emph {et~al.}(2017)\citenamefont
  {Hangleiter}, \citenamefont {Kliesch}, \citenamefont {Schwarz},\ and\
  \citenamefont {Eisert}}]{hangleiter2017}%
  \BibitemOpen
  \bibfield  {author} {\bibinfo {author} {\bibfnamefont {D.}~\bibnamefont
  {Hangleiter}}, \bibinfo {author} {\bibfnamefont {M.}~\bibnamefont {Kliesch}},
  \bibinfo {author} {\bibfnamefont {M.}~\bibnamefont {Schwarz}}, \ and\
  \bibinfo {author} {\bibfnamefont {J.}~\bibnamefont {Eisert}},\ }\href@noop {}
  {\bibfield  {journal} {\bibinfo  {journal} {Quantum Science and Technology}\
  }\textbf {\bibinfo {volume} {2}},\ \bibinfo {pages} {015004} (\bibinfo {year}
  {2017})}\BibitemShut {NoStop}%
\bibitem [{\citenamefont {Wiesner}(2017)}]{wiesner17}%
  \BibitemOpen
  \bibfield  {author} {\bibinfo {author} {\bibfnamefont {K.}~\bibnamefont
  {Wiesner}},\ }\href@noop {} {\bibfield  {journal} {\bibinfo  {journal} {arXiv
  preprint arXiv:1705.06768}\ } (\bibinfo {year} {2017})}\BibitemShut {NoStop}%
\bibitem [{\citenamefont {Calude}\ and\ \citenamefont
  {Calude}(2017)}]{calude17}%
  \BibitemOpen
  \bibfield  {author} {\bibinfo {author} {\bibfnamefont {C.~S.}\ \bibnamefont
  {Calude}}\ and\ \bibinfo {author} {\bibfnamefont {E.}~\bibnamefont
  {Calude}},\ }\href@noop {} {\bibfield  {journal} {\bibinfo  {journal} {arXiv
  preprint arXiv:1712.01356}\ } (\bibinfo {year} {2017})}\BibitemShut {NoStop}%
\bibitem [{\citenamefont {Neill}\ \emph {et~al.}(2017)\citenamefont {Neill},
  \citenamefont {Roushan}, \citenamefont {Kechedzhi}, \citenamefont {Boixo},
  \citenamefont {Isakov}, \citenamefont {Smelyanskiy}, \citenamefont {Barends},
  \citenamefont {Burkett}, \citenamefont {Chen}, \citenamefont {Chen} \emph
  {et~al.}}]{neill17}%
  \BibitemOpen
  \bibfield  {author} {\bibinfo {author} {\bibfnamefont {C.}~\bibnamefont
  {Neill}}, \bibinfo {author} {\bibfnamefont {P.}~\bibnamefont {Roushan}},
  \bibinfo {author} {\bibfnamefont {K.}~\bibnamefont {Kechedzhi}}, \bibinfo
  {author} {\bibfnamefont {S.}~\bibnamefont {Boixo}}, \bibinfo {author}
  {\bibfnamefont {S.}~\bibnamefont {Isakov}}, \bibinfo {author} {\bibfnamefont
  {V.}~\bibnamefont {Smelyanskiy}}, \bibinfo {author} {\bibfnamefont
  {R.}~\bibnamefont {Barends}}, \bibinfo {author} {\bibfnamefont
  {B.}~\bibnamefont {Burkett}}, \bibinfo {author} {\bibfnamefont
  {Y.}~\bibnamefont {Chen}}, \bibinfo {author} {\bibfnamefont {Z.}~\bibnamefont
  {Chen}},  \emph {et~al.},\ }\href@noop {} {\bibfield  {journal} {\bibinfo
  {journal} {arXiv preprint arXiv:1709.06678}\ } (\bibinfo {year}
  {2017})}\BibitemShut {NoStop}%
\bibitem [{\citenamefont {Miller}\ \emph {et~al.}(2017)\citenamefont {Miller},
  \citenamefont {Sanders},\ and\ \citenamefont {Miyake}}]{miller2017}%
  \BibitemOpen
  \bibfield  {author} {\bibinfo {author} {\bibfnamefont {J.}~\bibnamefont
  {Miller}}, \bibinfo {author} {\bibfnamefont {S.}~\bibnamefont {Sanders}}, \
  and\ \bibinfo {author} {\bibfnamefont {A.}~\bibnamefont {Miyake}},\
  }\href@noop {} {\bibfield  {journal} {\bibinfo  {journal} {arXiv preprint
  arXiv:1703.11002}\ } (\bibinfo {year} {2017})}\BibitemShut {NoStop}%
\bibitem [{\citenamefont {Bermejo-Vega}\ \emph {et~al.}(2017)\citenamefont
  {Bermejo-Vega}, \citenamefont {Hangleiter}, \citenamefont {Schwarz},
  \citenamefont {Raussendorf},\ and\ \citenamefont {Eisert}}]{bermejo2017}%
  \BibitemOpen
  \bibfield  {author} {\bibinfo {author} {\bibfnamefont {J.}~\bibnamefont
  {Bermejo-Vega}}, \bibinfo {author} {\bibfnamefont {D.}~\bibnamefont
  {Hangleiter}}, \bibinfo {author} {\bibfnamefont {M.}~\bibnamefont {Schwarz}},
  \bibinfo {author} {\bibfnamefont {R.}~\bibnamefont {Raussendorf}}, \ and\
  \bibinfo {author} {\bibfnamefont {J.}~\bibnamefont {Eisert}},\ }\href@noop {}
  {\bibfield  {journal} {\bibinfo  {journal} {arXiv preprint arXiv:1703.00466}\
  } (\bibinfo {year} {2017})}\BibitemShut {NoStop}%
\bibitem [{\citenamefont {Aaronson}\ and\ \citenamefont
  {Chen}(2016)}]{aaronson2016}%
  \BibitemOpen
  \bibfield  {author} {\bibinfo {author} {\bibfnamefont {S.}~\bibnamefont
  {Aaronson}}\ and\ \bibinfo {author} {\bibfnamefont {L.}~\bibnamefont
  {Chen}},\ }\href@noop {} {\bibfield  {journal} {\bibinfo  {journal} {arXiv
  preprint arXiv:1612.05903}\ } (\bibinfo {year} {2016})}\BibitemShut {NoStop}%
\bibitem [{\citenamefont {Bremner}\ \emph {et~al.}(2017)\citenamefont
  {Bremner}, \citenamefont {Montanaro},\ and\ \citenamefont
  {Shepherd}}]{bremner2016}%
  \BibitemOpen
  \bibfield  {author} {\bibinfo {author} {\bibfnamefont {M.~J.}\ \bibnamefont
  {Bremner}}, \bibinfo {author} {\bibfnamefont {A.}~\bibnamefont {Montanaro}},
  \ and\ \bibinfo {author} {\bibfnamefont {D.~J.}\ \bibnamefont {Shepherd}},\
  }\href {\doibase 10.22331/q-2017-04-25-8} {\bibfield  {journal} {\bibinfo
  {journal} {{Quantum}}\ }\textbf {\bibinfo {volume} {1}},\ \bibinfo {pages}
  {8} (\bibinfo {year} {2017})}\BibitemShut {NoStop}%
\bibitem [{\citenamefont {Boixo}\ \emph {et~al.}(2016)\citenamefont {Boixo},
  \citenamefont {Isakov}, \citenamefont {Smelyanskiy}, \citenamefont {Babbush},
  \citenamefont {Ding}, \citenamefont {Jiang}, \citenamefont {Martinis},\ and\
  \citenamefont {Neven}}]{boixo2016}%
  \BibitemOpen
  \bibfield  {author} {\bibinfo {author} {\bibfnamefont {S.}~\bibnamefont
  {Boixo}}, \bibinfo {author} {\bibfnamefont {S.~V.}\ \bibnamefont {Isakov}},
  \bibinfo {author} {\bibfnamefont {V.~N.}\ \bibnamefont {Smelyanskiy}},
  \bibinfo {author} {\bibfnamefont {R.}~\bibnamefont {Babbush}}, \bibinfo
  {author} {\bibfnamefont {N.}~\bibnamefont {Ding}}, \bibinfo {author}
  {\bibfnamefont {Z.}~\bibnamefont {Jiang}}, \bibinfo {author} {\bibfnamefont
  {J.~M.}\ \bibnamefont {Martinis}}, \ and\ \bibinfo {author} {\bibfnamefont
  {H.}~\bibnamefont {Neven}},\ }\href@noop {} {\bibfield  {journal} {\bibinfo
  {journal} {arXiv preprint arXiv:1608.00263}\ } (\bibinfo {year}
  {2016})}\BibitemShut {NoStop}%
\bibitem [{\citenamefont {Hanson}\ \emph {et~al.}(2007)\citenamefont {Hanson},
  \citenamefont {Kouwenhoven}, \citenamefont {Petta}, \citenamefont {Tarucha},\
  and\ \citenamefont {Vandersypen}}]{Hanson17}%
  \BibitemOpen
  \bibfield  {author} {\bibinfo {author} {\bibfnamefont {R.}~\bibnamefont
  {Hanson}}, \bibinfo {author} {\bibfnamefont {L.~P.}\ \bibnamefont
  {Kouwenhoven}}, \bibinfo {author} {\bibfnamefont {J.~R.}\ \bibnamefont
  {Petta}}, \bibinfo {author} {\bibfnamefont {S.}~\bibnamefont {Tarucha}}, \
  and\ \bibinfo {author} {\bibfnamefont {L.~M.~K.}\ \bibnamefont
  {Vandersypen}},\ }\href {\doibase 10.1103/RevModPhys.79.1217} {\bibfield
  {journal} {\bibinfo  {journal} {Rev. Mod. Phys.}\ }\textbf {\bibinfo {volume}
  {79}},\ \bibinfo {pages} {1217} (\bibinfo {year} {2007})}\BibitemShut
  {NoStop}%
\bibitem [{\citenamefont {Gray}\ \emph {et~al.}(2016)\citenamefont {Gray},
  \citenamefont {Bayat}, \citenamefont {Puddy}, \citenamefont {Smith},\ and\
  \citenamefont {Bose}}]{gray16}%
  \BibitemOpen
  \bibfield  {author} {\bibinfo {author} {\bibfnamefont {J.}~\bibnamefont
  {Gray}}, \bibinfo {author} {\bibfnamefont {A.}~\bibnamefont {Bayat}},
  \bibinfo {author} {\bibfnamefont {R.~K.}\ \bibnamefont {Puddy}}, \bibinfo
  {author} {\bibfnamefont {C.~G.}\ \bibnamefont {Smith}}, \ and\ \bibinfo
  {author} {\bibfnamefont {S.}~\bibnamefont {Bose}},\ }\href {\doibase
  10.1103/PhysRevB.94.195136} {\bibfield  {journal} {\bibinfo  {journal} {Phys.
  Rev. B}\ }\textbf {\bibinfo {volume} {94}},\ \bibinfo {pages} {195136}
  (\bibinfo {year} {2016})}\BibitemShut {NoStop}%
\bibitem [{\citenamefont {Porras}\ and\ \citenamefont
  {Cirac}(2004)}]{porras2004}%
  \BibitemOpen
  \bibfield  {author} {\bibinfo {author} {\bibfnamefont {D.}~\bibnamefont
  {Porras}}\ and\ \bibinfo {author} {\bibfnamefont {J.~I.}\ \bibnamefont
  {Cirac}},\ }\href@noop {} {\bibfield  {journal} {\bibinfo  {journal} {Phys.
  Rev. Lett.}\ }\textbf {\bibinfo {volume} {92}},\ \bibinfo {pages} {207901}
  (\bibinfo {year} {2004})}\BibitemShut {NoStop}%
\bibitem [{\citenamefont {Salath\'e}\ \emph {et~al.}(2015)\citenamefont
  {Salath\'e}, \citenamefont {Mondal}, \citenamefont {Oppliger}, \citenamefont
  {Heinsoo}, \citenamefont {Kurpiers}, \citenamefont
  {Poto\ifmmode~\check{c}\else \v{c}\fi{}nik}, \citenamefont {Mezzacapo},
  \citenamefont {Las~Heras}, \citenamefont {Lamata}, \citenamefont {Solano},
  \citenamefont {Filipp},\ and\ \citenamefont {Wallraff}}]{salath15}%
  \BibitemOpen
  \bibfield  {author} {\bibinfo {author} {\bibfnamefont {Y.}~\bibnamefont
  {Salath\'e}}, \bibinfo {author} {\bibfnamefont {M.}~\bibnamefont {Mondal}},
  \bibinfo {author} {\bibfnamefont {M.}~\bibnamefont {Oppliger}}, \bibinfo
  {author} {\bibfnamefont {J.}~\bibnamefont {Heinsoo}}, \bibinfo {author}
  {\bibfnamefont {P.}~\bibnamefont {Kurpiers}}, \bibinfo {author}
  {\bibfnamefont {A.}~\bibnamefont {Poto\ifmmode~\check{c}\else
  \v{c}\fi{}nik}}, \bibinfo {author} {\bibfnamefont {A.}~\bibnamefont
  {Mezzacapo}}, \bibinfo {author} {\bibfnamefont {U.}~\bibnamefont
  {Las~Heras}}, \bibinfo {author} {\bibfnamefont {L.}~\bibnamefont {Lamata}},
  \bibinfo {author} {\bibfnamefont {E.}~\bibnamefont {Solano}}, \bibinfo
  {author} {\bibfnamefont {S.}~\bibnamefont {Filipp}}, \ and\ \bibinfo {author}
  {\bibfnamefont {A.}~\bibnamefont {Wallraff}},\ }\href {\doibase
  10.1103/PhysRevX.5.021027} {\bibfield  {journal} {\bibinfo  {journal} {Phys.
  Rev. X}\ }\textbf {\bibinfo {volume} {5}},\ \bibinfo {pages} {021027}
  (\bibinfo {year} {2015})}\BibitemShut {NoStop}%
\bibitem [{\citenamefont {Gra{\ss}}\ and\ \citenamefont
  {Lewenstein}(2014)}]{grass2014}%
  \BibitemOpen
  \bibfield  {author} {\bibinfo {author} {\bibfnamefont {T.}~\bibnamefont
  {Gra{\ss}}}\ and\ \bibinfo {author} {\bibfnamefont {M.}~\bibnamefont
  {Lewenstein}},\ }\href@noop {} {\bibfield  {journal} {\bibinfo  {journal}
  {EPJ Quantum Technology}\ }\textbf {\bibinfo {volume} {1}},\ \bibinfo {pages}
  {8} (\bibinfo {year} {2014})}\BibitemShut {NoStop}%
\bibitem [{\citenamefont {Ma}\ \emph {et~al.}(2011)\citenamefont {Ma},
  \citenamefont {Dakic}, \citenamefont {Naylor}, \citenamefont {Zeilinger},\
  and\ \citenamefont {Walther}}]{ma2011}%
  \BibitemOpen
  \bibfield  {author} {\bibinfo {author} {\bibfnamefont {X.}~\bibnamefont
  {Ma}}, \bibinfo {author} {\bibfnamefont {B.}~\bibnamefont {Dakic}}, \bibinfo
  {author} {\bibfnamefont {W.}~\bibnamefont {Naylor}}, \bibinfo {author}
  {\bibfnamefont {A.}~\bibnamefont {Zeilinger}}, \ and\ \bibinfo {author}
  {\bibfnamefont {P.}~\bibnamefont {Walther}},\ }\href@noop {} {\bibfield
  {journal} {\bibinfo  {journal} {Nature Physics}\ }\textbf {\bibinfo {volume}
  {7}},\ \bibinfo {pages} {399} (\bibinfo {year} {2011})}\BibitemShut {NoStop}%
\bibitem [{\citenamefont {Christandl}\ \emph {et~al.}(2004)\citenamefont
  {Christandl}, \citenamefont {Datta}, \citenamefont {Ekert},\ and\
  \citenamefont {Landahl}}]{StateTransfer}%
  \BibitemOpen
  \bibfield  {author} {\bibinfo {author} {\bibfnamefont {M.}~\bibnamefont
  {Christandl}}, \bibinfo {author} {\bibfnamefont {N.}~\bibnamefont {Datta}},
  \bibinfo {author} {\bibfnamefont {A.}~\bibnamefont {Ekert}}, \ and\ \bibinfo
  {author} {\bibfnamefont {A.~J.}\ \bibnamefont {Landahl}},\ }\href {\doibase
  10.1103/PhysRevLett.92.187902} {\bibfield  {journal} {\bibinfo  {journal}
  {Phys. Rev. Lett.}\ }\textbf {\bibinfo {volume} {92}},\ \bibinfo {pages}
  {187902} (\bibinfo {year} {2004})}\BibitemShut {NoStop}%
\bibitem [{\citenamefont {Petta}\ \emph {et~al.}(2005)\citenamefont {Petta},
  \citenamefont {Johnson}, \citenamefont {Taylor}, \citenamefont {Laird},
  \citenamefont {Yacoby}, \citenamefont {Lukin}, \citenamefont {Marcus},
  \citenamefont {Hanson},\ and\ \citenamefont {Gossard}}]{petta05}%
  \BibitemOpen
  \bibfield  {author} {\bibinfo {author} {\bibfnamefont {J.~R.}\ \bibnamefont
  {Petta}}, \bibinfo {author} {\bibfnamefont {A.~C.}\ \bibnamefont {Johnson}},
  \bibinfo {author} {\bibfnamefont {J.~M.}\ \bibnamefont {Taylor}}, \bibinfo
  {author} {\bibfnamefont {E.~A.}\ \bibnamefont {Laird}}, \bibinfo {author}
  {\bibfnamefont {A.}~\bibnamefont {Yacoby}}, \bibinfo {author} {\bibfnamefont
  {M.~D.}\ \bibnamefont {Lukin}}, \bibinfo {author} {\bibfnamefont {C.~M.}\
  \bibnamefont {Marcus}}, \bibinfo {author} {\bibfnamefont {M.~P.}\
  \bibnamefont {Hanson}}, \ and\ \bibinfo {author} {\bibfnamefont {A.~C.}\
  \bibnamefont {Gossard}},\ }\href@noop {} {\bibfield  {journal} {\bibinfo
  {journal} {Science}\ }\textbf {\bibinfo {volume} {309}},\ \bibinfo {pages}
  {2180} (\bibinfo {year} {2005})}\BibitemShut {NoStop}%
\bibitem [{\citenamefont {Taylor}\ \emph {et~al.}(2007)\citenamefont {Taylor},
  \citenamefont {Petta}, \citenamefont {Johnson}, \citenamefont {Yacoby},
  \citenamefont {Marcus},\ and\ \citenamefont {Lukin}}]{taylor07}%
  \BibitemOpen
  \bibfield  {author} {\bibinfo {author} {\bibfnamefont {J.~M.}\ \bibnamefont
  {Taylor}}, \bibinfo {author} {\bibfnamefont {J.~R.}\ \bibnamefont {Petta}},
  \bibinfo {author} {\bibfnamefont {A.~C.}\ \bibnamefont {Johnson}}, \bibinfo
  {author} {\bibfnamefont {A.}~\bibnamefont {Yacoby}}, \bibinfo {author}
  {\bibfnamefont {C.~M.}\ \bibnamefont {Marcus}}, \ and\ \bibinfo {author}
  {\bibfnamefont {M.~D.}\ \bibnamefont {Lukin}},\ }\href {\doibase
  10.1103/PhysRevB.76.035315} {\bibfield  {journal} {\bibinfo  {journal} {Phys.
  Rev. B}\ }\textbf {\bibinfo {volume} {76}},\ \bibinfo {pages} {035315}
  (\bibinfo {year} {2007})}\BibitemShut {NoStop}%
\bibitem [{\citenamefont {Loss}\ and\ \citenamefont
  {DiVincenzo}(1998)}]{loss98}%
  \BibitemOpen
  \bibfield  {author} {\bibinfo {author} {\bibfnamefont {D.}~\bibnamefont
  {Loss}}\ and\ \bibinfo {author} {\bibfnamefont {D.~P.}\ \bibnamefont
  {DiVincenzo}},\ }\href {\doibase 10.1103/PhysRevA.57.120} {\bibfield
  {journal} {\bibinfo  {journal} {Phys. Rev. A}\ }\textbf {\bibinfo {volume}
  {57}},\ \bibinfo {pages} {120} (\bibinfo {year} {1998})}\BibitemShut
  {NoStop}%
\bibitem [{\citenamefont {Trotter}(1959)}]{trotter59}%
  \BibitemOpen
  \bibfield  {author} {\bibinfo {author} {\bibfnamefont {H.~F.}\ \bibnamefont
  {Trotter}},\ }\href@noop {} {\bibfield  {journal} {\bibinfo  {journal}
  {Proceedings of the American Mathematical Society}\ }\textbf {\bibinfo
  {volume} {10}},\ \bibinfo {pages} {545} (\bibinfo {year} {1959})}\BibitemShut
  {NoStop}%
\bibitem [{\citenamefont {Peres}(1984)}]{peres84}%
  \BibitemOpen
  \bibfield  {author} {\bibinfo {author} {\bibfnamefont {A.}~\bibnamefont
  {Peres}},\ }\href@noop {} {\bibfield  {journal} {\bibinfo  {journal}
  {Physical Review A}\ }\textbf {\bibinfo {volume} {30}},\ \bibinfo {pages}
  {1610} (\bibinfo {year} {1984})}\BibitemShut {NoStop}%
\bibitem [{\citenamefont {Jalabert}\ and\ \citenamefont
  {Pastawski}(2001)}]{jala01}%
  \BibitemOpen
  \bibfield  {author} {\bibinfo {author} {\bibfnamefont {R.~A.}\ \bibnamefont
  {Jalabert}}\ and\ \bibinfo {author} {\bibfnamefont {H.~M.}\ \bibnamefont
  {Pastawski}},\ }\href@noop {} {\bibfield  {journal} {\bibinfo  {journal}
  {Phys. Rev. Lett.}\ }\textbf {\bibinfo {volume} {86}},\ \bibinfo {pages}
  {2490} (\bibinfo {year} {2001})}\BibitemShut {NoStop}%
\end{thebibliography}%

\end{document}